# Laser-stimulated Synthesis of Large Nanostructured Fractal Silver Aggregates

A. K. Popov[♦,†,*], R. Tanke[†], J. Brummer[†], M. Loth[♦,§], R. Langlois[‡,§], A. Wruck[†,§], G. Taft[♦], and R. Schmitz[‡]

University of Wisconsin-Stevens Point, Stevens Point, Wisconsin 54481

**Abstract.** A Laser-stimulated synthesis of large silver nanoaggregates consisting of hundreds to thousands of nanoparticles is investigated. It is shown that the morphology of the synthesized nanostructures can be controlled with light and monitored via the evolution of the colloid absorption spectra. A solution method is demonstrated that enables production of a bulk amount of metal nanoaggregates, which are of paramount importance for subwavelength concentration and dramatic enhancement of the electromagnetically-induced processes at the nanoscale.

**Introduction.**

The high surface-to-volume ratios of nanoparticles lead to dramatic modifications in the material's properties. In particular, metal nanoparticles possess optical resonance which is determined not only by the electron plasma frequency, but also by the nanoparticle's size and shape. Such geometrical resonance is called "plasmonic" resonance. Particles behave in optical processes as artificial atoms and clusters of nanoparticles behave as artificial molecules. Similar to molecules, interaction between the constituent particles causes perturbation and shift of their plasmon resonances. The distance between the particles is inversely related to the shift observed. Due to the various distances between the particles inside the aggregate, formation of clusters leads to inhomogeneous broadening of the absorption resonance, which depends on distribution of distances between the neighbors inside the aggregate. A long tail stretched to the red side of the absorption spectrum indicates appearance of closely spaced particles. Metal nanostructured aggregates have enabled the concentration of optical electromagnetic radiation below the diffraction limits. As shown by Shalaev, Stockman and coworkers, this leads to such striking optical properties as giant enhancement and confinement of various optical processes at subwavelength scale (see, e.g.,[1-5] and references therein). Thus exciting avenues are opened for the creation of ultra small lasers and opto-electronic devices, for various applications in nanophotonics and biophotonics, for sensing of molecules, and for photomodification of biological cells.[1,4,5] Nanoaggregates can be further fixed in soft and solid matrices to form nanocomposites. Nanostructured silver metamaterials recently have attracted a great interest for creation of 'left-handed' media with a negative refractive index in the optical range.[6(a,b)] Successful experiments on realization of superlenses based on such plasmonic metamaterials have been reported.[6(c)]

Photo-assisted nanoengineering of nanostructured silver metamaterials can be incorporated into different synthesis methods.[7] Experiments where plasmon photoexcitation predominantly leads to conversion of a shape and change of the size of the isolated silver nanoparticles have been reported by Mirkin and coworkers (see[8] and references therein). Thus manipulating the morphological properties of the nanostructured metal composites by light presents an important goal of optical nanoengineering. Although various promising light-driven synthetic methods have been successfully realized to date (for review, see [9] and references therein), the detailed mechanisms that lead either to conversion of shape and size of isolated metal nanoparticles or, alternatively, to synthesis of the aggregates of small nanoparticles remain to be determined. Therefore, more experimental data are required in order to gain deeper insight into the role of the plasmon excitation and accompanying photochemical processes which are the basis for the selective optical nanoengineering. Herein, we report the experiments towards the photoinduced method for synthesizing large quantities of predominantly large fractal-type nanostructured silver aggregates in high yield in a bulk colloidal solution, which is stimulated and controlled by Ar-ion laser. The nanoengineering options are demonstrated, which depend upon the specific irradiation regimes and composition of the solution in the pre-irradiation stage.

**Basics of the synthetic technique.**

Neutral metal nanoparticles can be bound by the generalized van der Waals forces, which are very short-ranged down to about one nm. However, in many real situations, e.g., in colloids, nanoparticles are coated by other ions or polar molecules which prevent aggregation due to appearance of double electrical layers and repulsive forces. This leads to cessation of aggregation and to stabilization of the colloid. Light causes the escape of electrons from metals due to the photoeffect. This breaks the counterbalance of attracting and repulsive forces and may trigger further growth of the nanoparticles, changing their shape and forming nanoparticle clusters. Photostimulation of the aggregation rate and an up to $10^8$ fold accelerated formation of the nanostructured metal aggregates has been demonstrated.[2,3,10-11] Due to the enhancement of local fields by even small clusters, their presence in the preirradiation stage dramatically enhances photostimulation. For example, two-photon emission of electrons becomes possible even at the intensity of excitation on the level of sun light.[12] Whereas nonmonochromatic light enhances aggregation rate, photostimulation by strong laser irradiation is accompanied by photomodification of the synthesized aggregates.[13] Hence, the entire process is determined by the specific properties of a number of contributing processes such as intensity and wavelength dependence of plasmon excitation, photoeffect, photomodification and electrolytic properties of the solution, the shielding role of anions and polymers, etc. [2,3,9-11] Large aggregates usually have a rarefied often dendrite-type structure which possess the properties of statistical fractals and are characterized by the fractal dimensionality. Increased aggregation rates basically should lead to more rarefied structures possessing smaller fractal dimensionality, while more slow aggregation results in more dense structures. Thus photostimulation provides an extra means for manipulating the properties of the synthesized nanoaggregates through variation of the irradiation parameters and electrochemical properties of the colloids. Since the aggregation rate and properties of the synthesized aggregates depend upon the particular synthesis method, our objective is to investigate the outlined opportunities with the aid of the method described below.

The initial stage of the employed technique is based on reduction of cations in solution. Silver atoms begin to agglomerate and form nanoparticles. The reduction takes place in the presence of a stabilizing agent, which is adsorbed on the metal surface and stabilizes nanoparticles. Such stabilization provides colloidal nanoparticles in contrast with precipitation of bulk metal. Unlike [8], in our experiments, metal nanoparticles were synthesized in colloidal form by ethanol reduction of $AgNO_3$ and stabilized by the polymer PVP [(poly)vinyl pyrrolidone]. Initial aggregation in the colloid was controlled by amount of stabilizer used during the synthesis. Breaking stabilization and further formation of their aggregates was stimulated by continuous-wave green light from Argon-ion laser. The aggregation stages were monitored through the evolution of the absorption spectrum of the colloid. As outlined above, spectral broadening and stretching the spectrum to the red side indicated aggregation and appearance of the properties which enable subwavelength concentration of local optical fields. The size and structure of the aggregates was also directly characterized through their transmission electron microscope (TEM) images. We have also studied the regimes which alternatively provide change of the size and shape of nanoparticles and will be reported elsewhere.

**Experimental.**

In a typical preparation, a solution of 300.3 mg of PVP in 20mL of deionized water and 80mL of absolute ethanol was heated up to 77 °C in an Erlymeyer flask equipped with a stir bar. Then $AgNO_3$ (400.5 mg, 2.357 mmol) was added and heating continued at 77 °C for 20 min. The reaction was cooled and an aliquot was placed in a quartz cuvette of 1 x 1 x 4.5 cm size and irradiated by green $\lambda$ = 514.3 nm light of 30 mW Ar+ laser defocused to a beam waist of 2 cm. Another control aliquot was kept in the dark. At times the initially prepared mixture had orange tint and upon irradiation these samples lead to red shifted visible spectra as seen in **Figure 1** and aggregation in the TEM images as seen in **Figure 2**. At other times the initially prepared mixture was yellow and upon irradiation these samples did not show red shifted visible spectra color, see supporting materials for absorption spectra. Yellow color indicated the presence of isolated colloidal silver nanoparticles. Light orange tint indicated the presence of small nanoclusters resulted from heating. Appearance of nanoaggregates caused by irradiation of the sample led to drastic change of color, e.g., to orange-red and with prolonged irradiation to brown or gray. In another experiment, the ratio of PVP and $AgNO_3$ was slightly changed as 250 mg of PVP and 402.5 mg of $AgNO_3$. Change of this ratio and heating affects the appearance of initial small aggregates. It was determined that the process of photostimulation and its outcomes strongly depended on the amount of the stabilizing agent and heating.

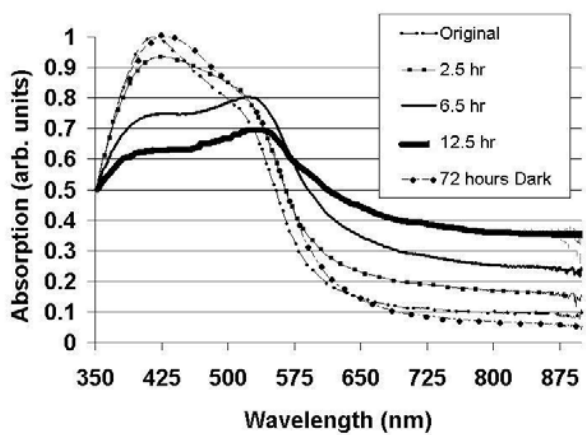

**Figure 1. Evolution of absorption spectrum of Ag colloid and appearance of absorption in red and near-IR ranges during irradiation by defocused Ar-ion laser.**

**Figure 1** presents evolution of colloid absorption spectra prepared as described above at 303.3/400.5 ratio of $PVP/AgNO_3$. Each spectrum is normalized to its maximum and then shifted to one and the same value at 350 nm. Striking features are the increase of absorption in green and yellow spectral ranges, and appearance of absorption in red and near-infrared wavelength ranges, which clearly indicate strong structural changes in the colloid.

According to [8], the appearance of extra red-shifted maximum suggests concomitant changes in shape and size of the constituent nanoparticles with sharp edges. The spectrum for the control sample kept in the dark practically does not change.

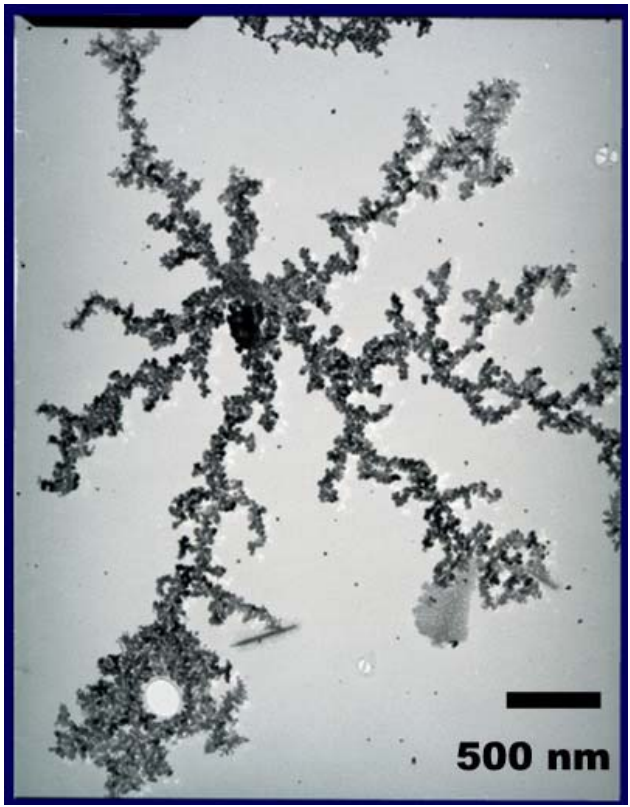

**Figure 2. "Octopus" - TEM image of the synthesized aggregate. The bar is equal to 250nm. The fractal's length is about 5,000 nm.**

Figures 2-4 show corresponding images of silver nanoparticles and their aggregates obtained with transmission electron microscopy (TEM). **Figure 2** presents TEM image of the aggregate found in the sample after about 5.5 hours of irradiation. It shows branching, dendrite-type rarefied structure which is characteristic for statistical fractals. Indeed, such kind of metal nanostructures enables concentration of light energy in the near-field zones and giant enhancement of nonlinear-optical processes at the nanoscale. **Figure 3** shows the TEM image of a part of the aggregate taken with higher resolution. These are two-dimensional projection of 3-D structure, which shows that the neighbor constituent nanoparticles are spaced at the distances much less than their sizes. The shape of the nanoparticles is not spherical, which explains the shift of the absorption maximum and the appearance of the additional peaks (see also higher-resolution images included in Supporting Material section). The image also depicts a structure typical for the statistical fractals. Characteristic TEM image of isolated nanoparticles and small aggregates taken from the sample kept in the dark is shown in **Figure 4**.

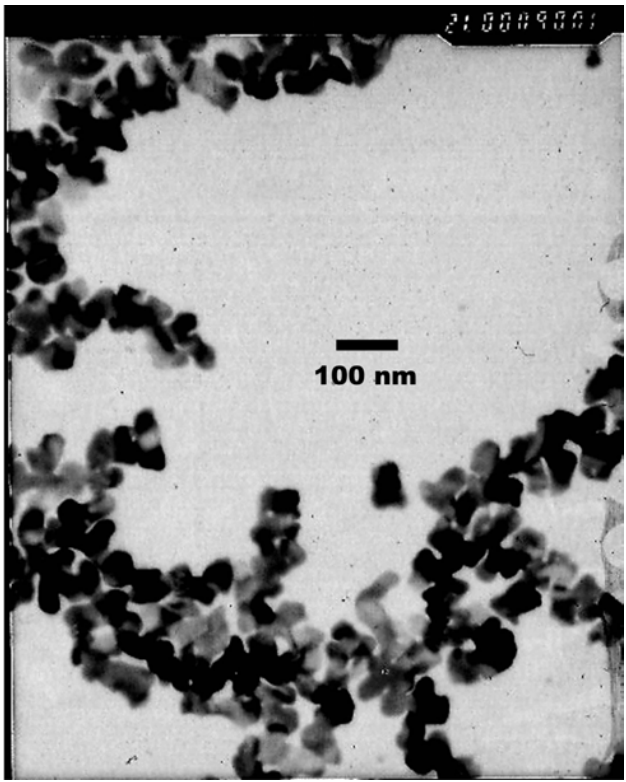

**Figure 3.** "Tentacle" - TEM image of a part of the aggregate taken with five fold higher magnification than Fig. 2.

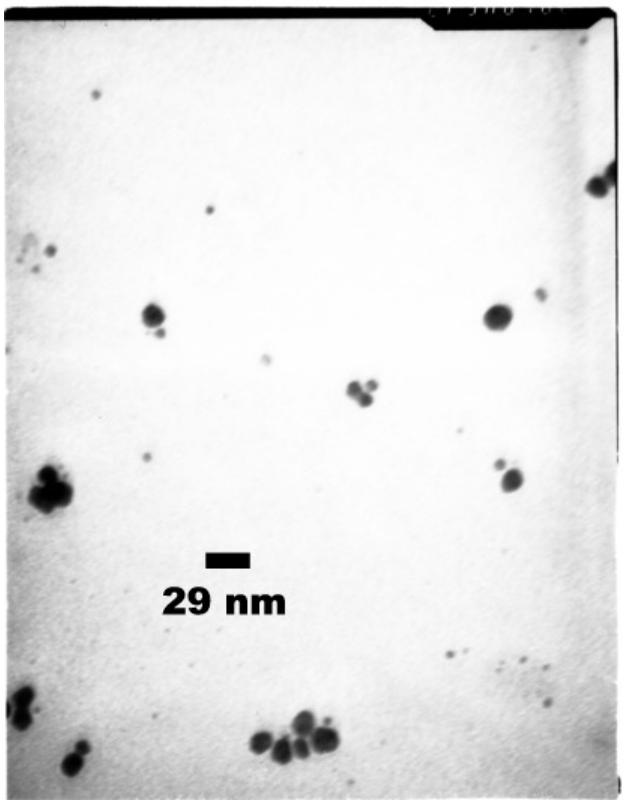

**Figure 4.** TEM images of the isolated silver nanoparticles and small aggregates in the control sample kept in the dark. The smallest monomers are about 9-11 nm.

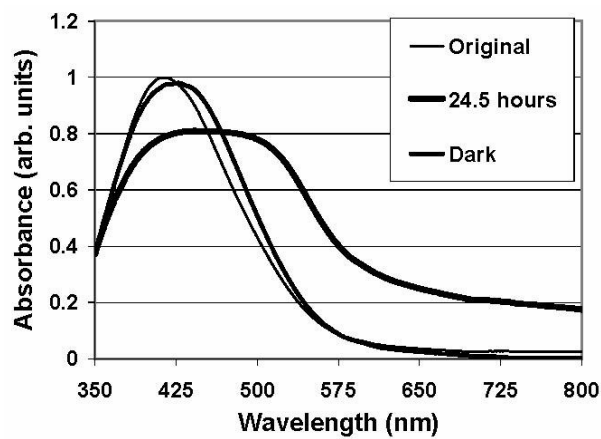

**Figure 5.** Evolution of the colloid's absorption spectra caused by irradiation of green $\lambda$ = 514.5 nm 28 mW Ar-ion laser light. Narrow peak – initial spectrum of the colloid, broad – after 24.5 hours of exposure to defocused Argon-ion laser irradiation, intermediate – control sample kept in the dark for 25.5 hours.

**Figures 5-8** present the experiment at slightly changed ratio 250/402.5 of PVP/AgNO$_3$ and other components remained the same. They display different absorption spectra and synthesized nanostructures. **Figure 5** presents modification of the absorption spectra. It shows similar to **Figure 1** basic feature: typical broadening, appearance of absorption in green and in the red tail in the absorption spectra in the process of photoinduced aggregation and practically unchanged spectrum for the control sample kept in the dark. However, different details in the spectra, such as absence of the secondary maximum, suggest different structures. It is confirmed by the TEM images presented in **Figures 6-8**.

**Figures 6-7** present the aggregates found in the sample after irradiation during 24.5 hours. (See also image with higher resolution presented in Supporting Information section.) The images depict a 3D structure consisting of hundreds of small particles with inclusion of small amount of bigger nanoparticles. The sizes of the constituent nanoparticles are about 1-20 nm. Most of them possess rounded edges.

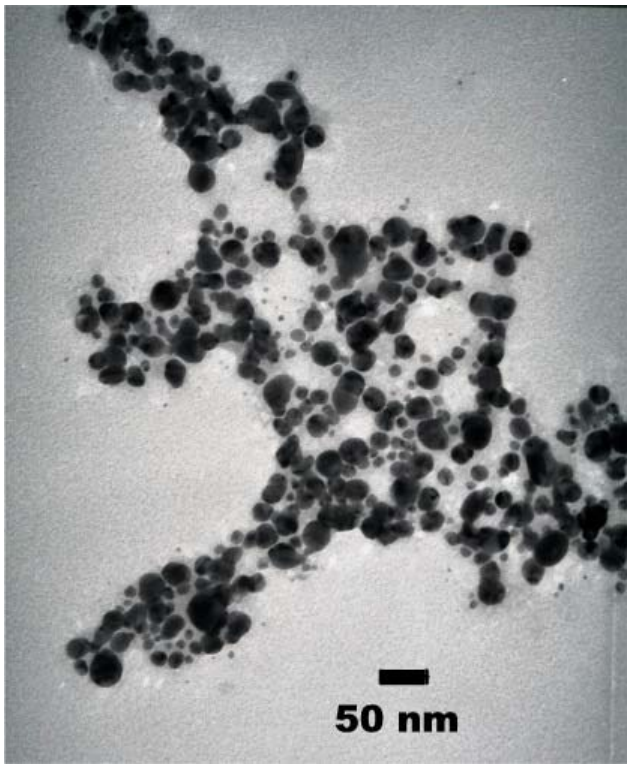

**Figure 6. "Charlie" - TEM image of the fragment of the aggregate of silver nanoparticles in the developed stage. The size of the fragment is about 1.8 µ.**

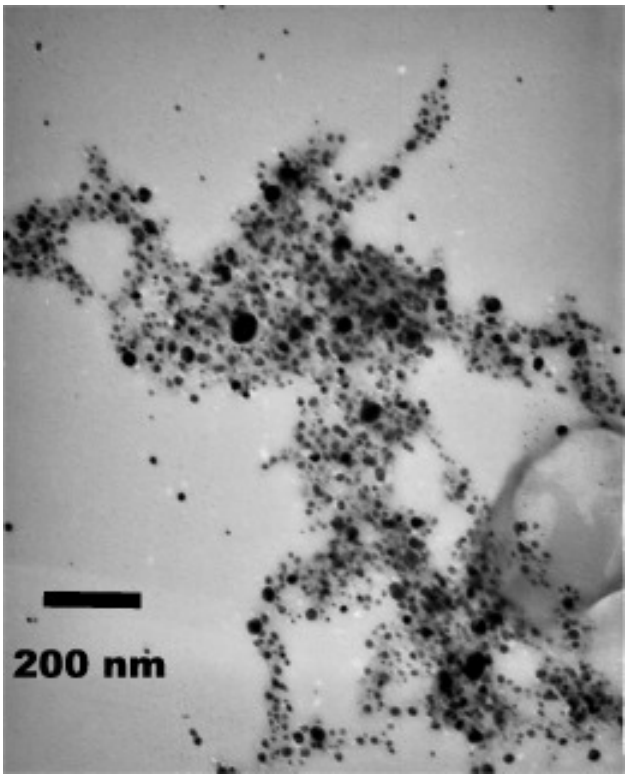

**Figure 7. "Warrior" -- TEM image of the nanostructure produced through photoinduced synthesis initiated by the Argon-ion laser irradiation.**

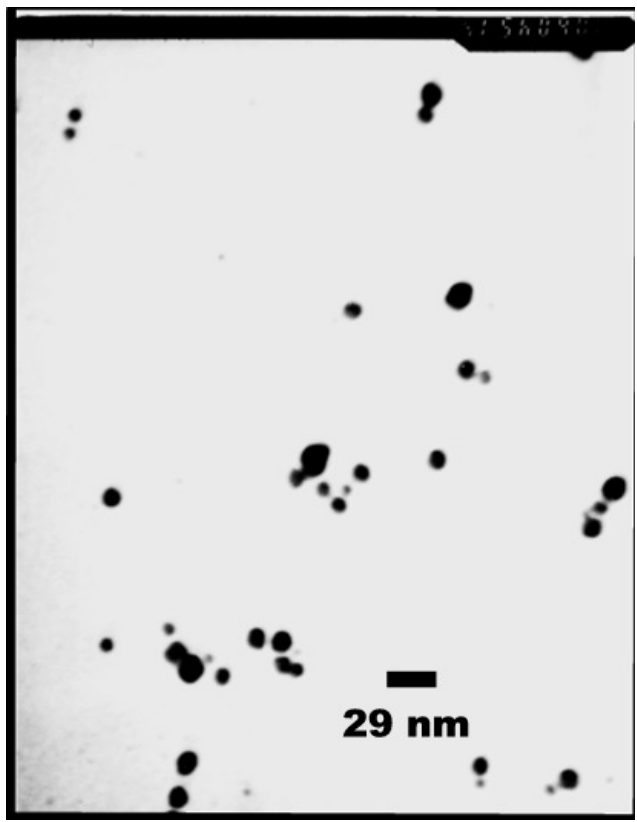

**Figure 8. TEM images of isolated nanoparticles and small aggregates in the sample before irradiation.**

Figure 8 displays isolated silver nanoparticles and small aggregates in the sample before irradiation by light. TEM images of nanoparticles and small aggregates taken from the sample kept in the dark are similar to that presented in Figures 4 and 8.

There is, however, a concern with sampling. Only a drop of the solution is adhered onto the carbon film that is used to obtain the TEM image, which may reduce the representation of the sample as a whole. In addition, the use of the electron beam in acquisition of the TEM image may alter the arrangement of the particles collected to some extend.

When investigating various combinations of initial stabilization and photostimulation regimes, which included different nonmonochromatic light sources, we have revealed such regimes that primarily stimulate a change of the size and shape of isolated nanoparticles rather then their aggregation. Along with the results described above, this provides the evidence of possibilities for application of the photostimulated synthesis in nanoengineerting.

**Conclusions.**

The feasibility of laser-stimulated synthesis of nanostructured metal aggregates consisting of hundreds to thousands of silver nanoparticles is investigated. Such aggregates have a great variety of important applications. It is proved that the nanoaggregates morphology can be controlled with laser and the specific features of such aggregation can be monitored via the evolution of the sample absorption spectrum. Photostimulated synthesis of plasmonic nanostructured aggregates and underlying photophysical and electrochemical processes are relatively scarcely studied, especially in the context of nanoengineering. Our experiments have provided the evidence that light produces photoeffect, which may substantially change the properties of the immediate environment of metal nanoparticles. Consequently, this may break stabilization and trigger the aggregation, which presents the feasibility for photostimulated aggregation and, hence, for manipulating properties of the plasmonic materials by light. The results essentially depend on the constituent nanoparticles' shape- and size-dependent plasmon resonance, on the content and thickness of the adsorbed layers and on the electrolytic (conducting) properties of the dispersive medium, on wavelength, intensity and temporal properties of the light source. A variety of the nanoengineering options depending on the specific synthesis regimes is shown to be feasible, which include conversion of shapes of isolated nanoparticles as well as size of the aggregates.

**Acknowledgments.**

AKP acknowledges the support of this work in part by the grant MDA972-03-1-0020 from DARPA and by the grant 00000496 from ARO. Electron Microscopy support for this project was provided by a Letters and Science Undergraduate Enrichment Initiative grant to R. J. Schmitz.

*Corresponding author. E-mail: apopov@uwsp.edu
[†]Department of Chemistry, [‡]Department of Biology, [♦]Department of Physics & Astronomy, [§]Undergraduate students